\documentclass[a4paper]{jpconf_pn} 

\usepackage{graphicx}
\usepackage{xcolor}
\usepackage{amsmath,amssymb,amsfonts,amsthm}
\usepackage{hyperref}

\def\osp{{\mathfrak{osp}}}

\def\sl{{\mathfrak{sl}}}
\def\os{{\mathfrak{o}}}
\def\su{{\mathfrak{su}}}
\def\sp{{\mathfrak{sp}}}
\def\bo{{\mathcal{O}}}
\def\ji{{\mathcal{J}}}

\begin{document}
\setcounter{page}{1} 
\title[The generalized Racah algebra as a commutant]{The generalized Racah algebra as a commutant}

\author{Julien Gaboriaud}
\ead{julien.gaboriaud@umontreal.ca}
\address{Centre de Recherches Math\'ematiques, Universit\'e de Montr\'eal, Montr\'eal (QC), Canada}

\author{Luc Vinet}
\ead{luc.vinet@umontreal.ca}
\address{Centre de Recherches Math\'ematiques, Universit\'e de Montr\'eal, Montr\'eal (QC), Canada}

\author{St\'ephane Vinet}
\ead{stephanevinet@uchicago.edu}
\address{The College, The University of Chicago, 5801 S. Ellis Ave, Chicago, IL 60637, USA}

\author{Alexei Zhedanov}
\ead{zhedanov@yahoo.com}
\address{Department of Mathematics, Renmin University of China, Beijing 100872, China}

\begin{abstract}
The Racah algebra $R(n)$ of rank $(n-2)$ is obtained as the commutant of the \mbox{$\mathfrak{o}(2)^{\oplus n}$} subalgebra of $\mathfrak{o}(2n)$ in oscillator representations of the universal algebra of $\mathfrak{o}(2n)$. This result is shown to be related in a Howe duality context to the definition of $R(n)$ as the algebra of Casimir operators arising in recouplings of $n$ copies of $\mathfrak{su}(1,1)$. These observations provide a natural framework to carry out the derivation by dimensional reduction of the generic superintegrable model on the $(n-1)$ sphere which is invariant under $R(n)$.
\end{abstract}

\section{Introduction}\label{sec_intro}
The Racah algebra $R(3)$ of rank $1$ \cite{Granovskii1988,Genest2014a} encodes the bispectrality properties of the Racah polynomials \cite{Koekoek2010} and is the symmetry algebra of the generic superintegrable model on the \mbox{$2$-sphere} with Hamiltonian $H$ given by \cite{Kalnins2007}
\begin{equation}\label{eq_H}
 H=\sum_{1\leq i<j\leq 3}{\ji_{ij}}^2+\sum_{i=1}^{3}\frac{a_i}{{x_i}^2} 
\end{equation}
where 
\begin{equation}\label{eq_Jk1}
 \ji_{ij}=x_i \frac{\partial}{\partial x_j}-x_j \frac{\partial}{\partial x_i}, \quad\qquad {x_1}^2+{x_2}^2+{x_3}^2=1
\end{equation}
and $a_1$, $a_2$, $a_3$ are parameters. For a review the reader is referred to \cite{Genest2014}. Of particular relevance is the fact that $R(3)$ was seen to be  the commutant in $\mathcal{U}(\su(1,1)^{\otimes{3}})$ of the embedding of $\su(1,1)$ in the three-fold tensor product of this algebra with itself, or in other words, that it is generated by the invariant operators arising in this Racah problem. This observation provided a way to generalize $R(3)$ to Racah algebras of arbitrary rank $(n-2)$ \cite{DeBie2017} by extending the picture to $n$ factors and identifying the structure relations between the various Casimir operators arising in the possible recouplings. It follows that $R(n)$ thus defined is the symmetry algebra of the superintegrable model on the $(n-1)$-sphere obtained by straighforwardly extending to $n$ variables the model on $S^2$ defined above.

We have found recently \cite{Gaboriaud2018} that $R(3)$ can be realized as the commutant of the subalgebra $\os(2)\oplus\os(2)\oplus\os(2)\subset\os(6)$ in oscillator representations of the enveloping algebra of $\os(6)$. We further observed that this description of $R(3)$ could be related to the one associated to the Racah problem for $\su(1,1)$ through the Howe duality corresponding to the pair $(\os(6), \su(1,1))$. This provided a natural background for obtaining the superintegrable Hamiltonian \eref{eq_H} with $R(3)$ as symmetry algebra, under the dimensional reduction of a six-dimensional harmonic oscillator problem. We here wish to indicate how these results extend for $R(n)$, that is, for arbitrary ranks and dimensions.

The paper is structured as follows. In Section \ref{sec_gen_racah_su11}, we review how the Racah algebra $R(n)$ is defined as the algebra of the Casimir operators in the $n$-fold tensor product of $\su(1,1)$ Lie algebras. The structure relations satisfied by these Casimirs are provided. In Section \ref{sec_racah_o2n}, we show that the generators of the commutant of the $\os(2)^{\oplus n}$ subalgebra of $\os(2n)$ satisfy the defining relations of $R(n)$. In Section \ref{sec_racah_howe}, we invoke Howe duality to explain how the pairings between representations of $\os(2n)$ and those of $\su(1,1)$ underpin the connection between the tensorial and the commutant pictures of $R(n)$. How the $R(n)$-invariant superintegrable model on $S^{(n-1)}$ is obtained from an $n$-dimensional harmonic oscillator by modding out the action of the $n$-torus group is described in Section \ref{sec_racah_generic_model_Sn} and conclusions form Section \ref{sec_conclusion}.

\section{The generalized Racah algebra and tensor products of $\mathfrak{su}(1,1)$}\label{sec_gen_racah_su11}
Let us recall how the generalized Racah algebra $R(n)$ is defined from the $n$-fold tensor product of $\su(1,1)$. The $\su(1,1)$ algebra has $3$ generators, $J_0$, $J_\pm$ obeying the commutation relations
\begin{align}
 [J_0,J_\pm]=\pm J_\pm,  \qquad [J_+,J_-]=-2J_0.
\end{align}
The Casimir element is given by
\begin{align}
 C={J_0}^{2}-J_+J_--J_0.
\end{align}
Let $[n]=\{1,2,\dots,n\}$ denote the set of the $n$ first integers and consider the tensor product $\su(1,1)^{\otimes n}$. Coproduct embeddings of $\su(1,1)$ in $\su(1,1)^{\otimes n}$ are labelled by subsets $A \subset [n]$ with the generators mapped to
\begin{align}
 J^{A}=\sum_{i\in A}J^{i}
\end{align}
and where the superindex denotes on which factor of $\su(1,1)^{\otimes n}$ the operator $J^{i}$ is acting. Correspondingly, the Casimirs are sent to
\begin{align}
 C^{A}=\left(J_0^{A}\right)^{2}-J_+^{A}J_-^{A}-J_0^{A}.
\end{align}
The generalized Racah algebra $R(n)$ is taken to be the algebra generated by all these intermediate Casimirs $C^{A}$ since this is the case for $R(3)$.

It is important to note that not all intermediate Casimirs $C^{A}$ are independent; indeed one has
\begin{align}\label{eq_hr_indCas}
C^A=\sum_{\left\{i,j\right\}\subset A} C^{ij}-\left(|A|-2\right)\sum_{i \in A} C^i.
\end{align}
where $|A|$ stands for the cardinality of $A$. In order to characterize $R(n)$, given that the elements $C^i$ are central, it therefore suffices to provide all the iterated commutators between the $C^{ij}$'s with $i\neq j$ until closure is achieved. This has been carried out in \cite{DeBie2017}. It is convenient to introduce $P^{ij}$ and $F^{ijk}$:
\begin{align}\label{eq_genRn}
P^{ij} = C^{ij}-C^i-C^j,\qquad F^{ijk} = \frac{1}{2}[P^{ij},P^{jk}].
\end{align}
The defining relations of the Racah algebra $R(n)$ then read
\begin{subequations}\label{eq_Rr}
\begin{align}
 [P^{ij},P^{jk}] &= 2 F^{ijk},\label{eq_Rn1}\\
 [P^{jk},F^{ijk}] &= P^{ik}P^{jk}-P^{jk}P^{ij}+2P^{ik}C^j-2P^{ij}C^k,\label{eq_Rn2}\\
 [P^{kl},F^{ijk}] &= P^{ik}P^{jl}-P^{il}P^{jk},\label{eq_Rn3}\\
 [F^{ijk},F^{jkl}] &= F^{jkl}P^{ij}-F^{ikl}\big(P^{jk}+2C^j\big)-F^{ijk}P^{jl},\label{eq_Rn4}\\
 [F^{ijk},F^{klm}] &= F^{ilm}P^{jk}-P^{ik}F^{jlm},\label{eq_Rn5}
\end{align} 
\end{subequations}
where $i,j,k,l,m\in[n]$ are all different.

In the rank 1 case, \eref{eq_Rn3}, \eref{eq_Rn4} and \eref{eq_Rn5} are redundant and the standard Racah algera $R(3)$ is fully described by \eref{eq_Rn1} and \eref{eq_Rn2}. Note that the presentation that results from the specialization of these equations to $n=3$ is the equitable one. The relation between this presentation and the standard one used in \cite{Gaboriaud2018} is given explicitly in \cite{Genest2013}. The rank $2$ Racah algebra (which has been studied in detail in \cite{Post2015}) only requires \eref{eq_Rn1}-\eref{eq_Rn3} to be characterized, while the relations \eref{eq_Rn4} and \eref{eq_Rn5} have to be added in order to define Racah algebras of rank $3$ or higher.

\section{The generalized Racah algebra and $\mathfrak{o}(2n)$}\label{sec_racah_o2n}
Let us now indicate how the relations \eref{eq_Rr} given above are satisfied by the generators in\, $\mathcal{U}(\os(2n))$ of the commutant of $n$ copies of $\os(2)$ sitting in $\os(2n)$. The algebra $\os(2n)$ has $n(2n-1)$ generators $L_{\mu\nu}=-L_{\nu\mu}$,~ $\mu,\nu=1,...,2n$ obeying 
\begin{align}\label{eq_o2n_comm}
 [L_{\mu\nu},L_{\rho\sigma}]=\delta_{\nu\rho}L_{\mu\sigma}-\delta_{\nu\sigma}L_{\mu\rho}-\delta_{\mu\rho}L_{\nu\sigma}+\delta_{\mu\sigma}L_{\nu\rho}
\end{align}
and possesses the following quadratic Casimir:
\begin{align}\label{eq_Cas_o2n}
 \mathcal{C}=\hspace{-0.4em}\sum_{1\leq\mu<\nu\leq n}\hspace{-0.4em}L_{\mu\nu}^2.
\end{align}
We will use the realization
\begin{align}\label{eq_realization_L}
 L_{\mu\nu}=\xi_\mu\frac{\partial}{\partial\xi_\nu}-\xi_\nu\frac{\partial}{\partial\xi_\mu},\qquad \mu\neq \nu,\qquad \mu,\nu=1,\dots,2n.
\end{align}
Pick the \mbox{$\os(2)^{\oplus n}$} 
subalgebra of $\os(2n)$ generated by the commutative set \mbox{$\{L_{12},\,L_{34},\,\dots,\,L_{2n-1,2n}\}$}. We want to focus on the commutant in $\mathcal{U}(\os(2n))$ of this Abelian algebra.

It is easy to see that the set of invariants $\{G^{i},K^{ij}\}_{1\leq i<j\leq n}$,
\begin{align}
 G^{i}&=L_{2i-1,2i}^{2},\label{eq_inv1}\\
 K^{ij}&=L_{2i-1,2i}^2+L_{2i-1,2j-1}^2+L_{2i-1,2j}^2+L_{2i,2j-1}^2+L_{2i,2j}^2+L_{2j-1,2j}^2,\label{eq_inv2}
\end{align}
is sufficient to generate this commutant and it happens to be the generalized Racah algebra. Indeed, with the following redefinitions
\begin{align}
 C^{i}&=-\frac{1}{4}G^{i}+\frac{1}{4},\\
 C^{ij}&=-\frac{1}{4}K^{ij},\\
 P^{ij}&=-\frac{1}{4}K^{ij}+\frac{1}{4}\left(G^{i}+G^{j}\right)+\frac{1}{2},\\
 F^{ijk}&=\frac{1}{32}[K^{ij},K^{jk}],
\end{align}
a long but straightforward calculation in the realization \eref{eq_realization_L} shows that the defining relations \eref{eq_Rr} of the algebra $R(n)$ are obeyed.

\section{The $\mathfrak{su}(1,1)$ and $\mathfrak{o}(2n)$ descriptions of $R(n)$ and Howe duality}\label{sec_racah_howe}
In the last two sections we indicated that the generalized Racah algebra $R(n)$ is the commutant of $\su(1,1)$ in\, $\mathcal{U}\big(\su(1,1)^{\otimes n}\big)$ and of \mbox{$\os(2)^{\oplus n}$} in oscillator representations of\, $\mathcal{U}(\os(2n))$. The connection between these two descriptions is rooted in Howe duality.

It is known \cite{Howe1987,Howe1989,Howe1989a,Rowe2012} that $\os(2n)$ and $\sp(2)$ form a dual pair in $\sp(4n)$, with these two subalgebras being their mutual commutants. This implies that $\os(2n)$ and $\sp(2)\simeq \su(1,1)$ have dual actions on the Hilbert space of $2n$ oscillator states. That means that their irreducible representations can be paired and this can be done through the Casimirs in the following way.

Consider the $2n$ copies of the metaplectic realization of $\sp(2)$:
\begin{align}\label{eq_su11J}
 J_{+}^{(\mu)}=\frac{1}{2}\xi_\mu^2,\qquad\ J_{-}^{(\mu)}=\frac{1}{2}\frac{\partial^2}{\partial\xi_\mu^2},\qquad\ J_0^{(\mu)}=\frac{1}{2}\left(\frac{1}{2}+\xi_\mu\frac{\partial}{\partial\xi_\mu}\right),\qquad\ \mu=1,2,\dots2n.
\end{align}
We first add these $2n$ representations by coupling them pairwise
\begin{align}
 J^{(\mu;\nu)}=J^{(\mu)}+J^{(\nu)}.
\end{align}
In what follows, we will always assume that the pairs denoted $(\mu;\nu)$ are such that \mbox{$(\mu;\nu)=(2i-1;2i)$},~ \mbox{$i=1,\dots,n$}.
Now take $A\subset[n]$ to be any subset that is the union of $N$ such pairs:
\begin{align}
 A=\bigcup_{i=1}^{N}\{\mu_i;\nu_i\},
\end{align}
with $|A|=2N$ and $1\leq N\leq n$. The $\su(1,1)$ realization associated to such a subset $A$ reads
\begin{align}
 J_+^{A}=\frac{1}{2}\sum_{\mu\in A}\xi_\mu^{2},\qquad
 J_-^{A}=\frac{1}{2}\sum_{\mu\in A}\frac{\partial^{2}}{\partial\xi_\mu^{2}},\qquad
 J_0^{A}=\frac{1}{2}\left(\frac{|A|}{2}+\sum_{\mu\in A}\xi_\mu\frac{\partial}{\partial \xi_\mu}\right).
\end{align}
It is then straightforward to show that the Casimir for an embedding labelled by the subset $A$ is given by
\begin{align}
 C^{A}&=\left(J_0^{A}\right)^{2}-J_+^{A}J_-^{A}-J_0^{A}=\frac{|A|(|A|-4)}{16}-\sum_{\substack{\mu<\nu\\ \mu,\nu\in A}}\frac{\left(L_{\mu\nu}\right)^{2}}{4}.
\end{align}
As already noted, not all $C^{A}$'s are independent. The translation of \eref{eq_hr_indCas} shows that all $C^{A}$'s can be rewritten as
\begin{align}
  C^{A}=\hspace{-1.em}\sum_{\substack{(\mu;\nu),(\rho;\sigma)\in A \\ \mu<\nu<\rho<\sigma}}\hspace{-1.em}C^{(\mu;\nu)(\rho;\sigma)}-\frac{|A|-4}{2}\sum_{(\mu;\nu)\in A}C^{(\mu;\nu)}
\end{align}
with
\begin{align}\label{eq_ci_cij}
 C^{(\mu;\nu)}=-\frac{1}{4}\left(L_{\mu\nu}^{2}+1\right),\qquad\quad C^{(\mu;\nu)(\rho;\sigma)}=-\frac{1}{4}\left(L_{\mu\nu}^{2}+L_{\mu\rho}^{2}+L_{\mu\sigma}^{2}+L_{\nu\rho}^{2}+L_{\nu\sigma}^{2}+L_{\rho\sigma}^{2}\right).
\end{align}
This shows  that all higher order Casimirs can be reexpressed in terms of those of lowest orders. 
%

We thus observe that the intermediate $\sp(2)$ Casimirs correspond (up to an affine transformation) to the generators of the commutant of $\{L_{1,2},\,\dots,\,L_{2n-1,2n}\}$ in $\mathcal{U}(\os(2n))$. We know from \Sref{sec_racah_o2n}, that the intermediate $\sp(2)$ Casimirs realize the commutation relations of the generalized Racah algebra. This will hence be the case also for the commutant generators and we have here our duality connection.

\section{The generalized Racah algebra and the generic superintegrable model on $S^{n-1}$}\label{sec_racah_generic_model_Sn}
We can now complete the picture by performing the dimensional reduction from $\mathbb R^{2n}$ to $\mathbb R^+  \times S^{n-1}$ to obtain the generic superintegrable model with Hamiltonian $H$ (introduced in \Sref{sec_intro}) and to recover its symmetries. Starting from the oscillator representation \eref{eq_su11J}, make the following change of variables: 
\begin{align}
\begin{aligned}{}
 \xi_{2i-1}&=x_i\cos{\theta_i},\\
 \xi_{2i}&=x_i\sin{\theta_i},
\end{aligned}
 \qquad L_{2i-1,2i}=\xi_{2i-1}\frac{\partial}{\partial\xi_{2i}}-\xi_{2i}\frac{\partial}{\partial\xi_{2i-1}}=\frac{\partial}{\partial\theta_i}, \qquad i=1,\dots,n.
\end{align}
Eliminate the ignorable $\theta_i$'s by separating these variables and setting $L_{2i-1,2i}^2\sim k_i^2$. After performing the gauge transformation $\bo\mapsto\widetilde{\bo}=x_i^{1/2}\,\,\bo\,\,x_i^{-1/2}$ one obtains the reduced system
\begin{align}
 \hspace{-1em}\widetilde{J_+}^{(2i-1,2i)}=\frac{1}{2}x_i^2,\qquad \widetilde{J_-}^{(2i-1,2i)}=\frac{1}{2}\left(\frac{\partial^2}{\partial x_i^2}+\frac{a_i}{x_i^2}\right),\qquad\widetilde{J_0}^{(2i-1,2i)}=\frac{1}{2}\left(x_i\frac{\partial}{\partial x_i}+\frac{1}{2}\right),
\end{align}
with $a_i=k_i^2+\frac{1}{4}$. Defining $\widetilde{J}^{i}\equiv\widetilde{J}^{(2i-1,2i)}$, the reduced Casimirs
\begin{align}
 \widetilde{C}^{i}&=\left(\widetilde{J_0}^{i}\right)^2-\widetilde{J_+}^{i}\widetilde{J_-}^{i}-\widetilde{J_0}^{i},\\
 \widetilde{C}^{ij}&=\left(\widetilde{J_0}^{i}+\widetilde{J_0}^{j}\right)^2-\left(\widetilde{J_+}^{i}+\widetilde{J_+}^{j}\right)\left(\widetilde{J_-}^{i}+\widetilde{J_-}^{j}\right)-\left(\widetilde{J_0}^{i}+\widetilde{J_0}^{j}\right),
\end{align}
are easily computed and have the following expressions:
\begin{align}
\begin{aligned}{}
 \widetilde{C}^{i}&=-\frac{1}{4}\left(a_i+\frac{3}{4}\right),\\
 \widetilde{C}^{ij}&=-\frac{1}{4}\left[{\ji_{ij}}^2+a_i\frac{x_j^2}{x_i^2}+a_j\frac{x_i^2}{x_j^2}+a_i+a_j+1\right],
\end{aligned}
\qquad \ji_{ij}=x_i\frac{\partial}{\partial x_j}-x_j\frac{\partial}{\partial x_i},\qquad i<j.
\end{align}
Using the fact that $\widetilde{J}^{[n]}=\sum_{i=1}^{n}\widetilde{J}^{i}$, the total Casimir $\widetilde{C}^{[n]}$ is obtained:
\begin{align}
 \widetilde{C}^{[n]}=-\frac{1}{4}\hspace{-0.1em}\sum_{1\leq i<j\leq n}\hspace{-0.8em}{\ji_{ij}}^{2}-\frac{1}{4}\left(\sum_{i=1}^{n}x_i^{2}\right)\sum_{j=1}^{n}\frac{a_j}{x_{j}^{2}}+\frac{n(n-4)}{16},
\end{align}
and assuming \mbox{$\sum_{i=1}^{n}{x_i}^2=1$}, one thereby obtains the Hamiltonian of the generic model on $S^{n-1}$ (up to an affine transformation).
The basic intermediate Casimirs are essentially the conserved quantities:
\begin{align}
 Q_{ij}={\ji_{ij}}^{2}+a_i\frac{x_j^2}{x_i^2}+a_j\frac{x_i^2}{x_j^2},\qquad\quad 1\leq i<j\leq n
\end{align}
and they generate $R(n)$ which is hence the symmetry algebra of the superintegrable model on the $(n-1)$ sphere. (Note that the $Q_{ij}$'s are affinely related to the $P^{ij}$'s in the relations \eref{eq_genRn}.)

\section{Conclusion}\label{sec_conclusion}
Summing up, we have shown that the generalized Racah algebra $R(n)$ can be defined as the commutant of the $\os(2)^{\oplus n}$ subalgebra of $\os(2n)$ in oscillator representations of $\mathcal{U}(\os(2n))$. This offers an alternative to the definition of $R(n)$ as the algebra of the intermediate Casimirs associated to the $\su(1,1)$ embeddings in $\su(1,1)^{\otimes n}$. We have related these two pictures in the context of Howe duality and obtained the generic $R(n)$-invariant superintegrable model on $S^{n-1}$ through the dimensional reduction scheme stemming from the analysis. This has provided a generalization to arbitrary ranks and dimensions of the study carried in \cite{Gaboriaud2018} for the standard Racah algebra.

We wish to remark that since $\os(nd)$ and $\sp(2)$ form a dual pair in $\sp(2nd)$, it is also  possible to realize the generalized Racah algebra as the commutant of the $\os(d)^{\oplus n}$ subalgebra of $\os(nd)$. We have concentrated on the case $d=2$ because it offers the simplest situation that allows to obtain the superintegrable system on $S^{n-1}$ by dimensional reduction.


In the near future, we plan on exploring similarly the Askey-Wilson (AW) and the Bannai-Ito (BI) algebras which share features with the Racah algebra since both encode the bispectrality properties of the eponym polynomials and appear through tensor products of $\mathcal{U}_q(\sl(2))$ \cite{Granovskii1993} and  $\osp(1|2)\simeq\sl_{-1}(2)$ \cite{Genest2014b} respectively. Moreover, the BI algebra is the symmetry algebra of a superintegrable model on the sphere involving reflection operators \cite{Genest2014c} as well as a Dirac-Dunkl equation in $\mathbb R^3$ \cite{DeBie2017}. It would be of interest to build on the work of the present paper to obtain a Howe duality setting for the interpretation of the AW and BI algebras as commutants; moreover extensions along the lines of this paper would shed interesting light on the higher rank versions of these algebras.

\ack
JG holds an Alexander-Graham-Bell scholarship from the Natural Science and Engineering Research Council (NSERC) of Canada. 
LV gratefully acknowledges his support from NSERC. 
Also SV enjoys a Neubauer No Barriers scholarship at the University of Chicago and benefitted from a Metcalf internship.
The work of AZ is supported by the National Foundation of China (Grant No. 11771015).\\

\section*{References}
\bibliographystyle{iopart-num}
\bibliography{citations}

\end{document}